\documentclass[prb,twocolumn,showpacs,superscriptaddress,preprintnumbers,amsmath,amssymb]{revtex4}

\usepackage{graphicx}%

\begin{document}

\title{Doping dependence of the Fermi surface in (Bi,Pb)$_2$Sr$_2$CaCu$_2$O$_{8+\delta}$}

\date{\today} 

\author{A. A. Kordyuk}
\affiliation{Institute for Solid State and Materials Research Dresden, P.O.Box 270016, D-01171 Dresden, Germany}
\affiliation{Institute of Metal Physics of National Academy of Sciencies of Ukraine, 03142 Kyiv, Ukraine}

\author{S. V. Borisenko}
\affiliation{Institute for Solid State and Materials Research Dresden, P.O.Box 270016, D-01171 Dresden, Germany}

\author{M. S. Golden}
\altaffiliation[New address: ]{Van der Waals-Zeeman Institute, University of Amsterdam, Valckenierstraat 65, NL 1018 XE Amsterdam, The Netherlands.}
\affiliation{Institute for Solid State and Materials Research Dresden, P.O.Box 270016, D-01171 Dresden, Germany}

\author{S. Legner}
\author{K.~A.~Nenkov}
\author{M. Knupfer}
\author{J. Fink}
\affiliation{Institute for Solid State and Materials Research Dresden, P.O.Box 270016, D-01171 Dresden, Germany}

\author{H. Berger}
\affiliation{Institut de Physique Appliqu\'ee, Ecole Politechnique F\'ederale de Lausanne, CH-1015 Lausanne, Switzerland}

\author{L. Forr\'{o}}
\affiliation{DP/IGA, Ecole Politechnique F\'ederale de Lausanne, CH-1015 Lausanne, Switzerland}

\author{R. Follath}
\affiliation{BESSY GmbH, Albert-Einstein-Strasse 15, 12489 Berlin, Germany}

\begin{abstract}
A detailed and systematic ARPES investigation of the doping-dependence of the normal state Fermi surface (FS) of modulation-free (Pb,Bi)-2212 is presented. The FS does not change in topology away from hole-like at any stage. The FS area does not follow the usual curve describing $T_c$ vs $x$ for the hole doped cuprates, but is down-shifted in doping by ca.~0.05 holes per Cu site, indicating the consequences of a significant bilayer splitting of the FS across the whole doping range. The strong $k$-dependence of the FS width is shown to be doping \textit{independent}. The relative strength of the shadow FS has a doping dependence mirroring that of $T_c$.
\end{abstract} 


\preprint{}

\maketitle

The shape and topology of the Fermi surface (FS) of the high temperature superconductors (HTSC), and in particular of the Bi$_2$Sr$_2$CaCu$_2$O$_{8+\delta}$ (Bi-2212)-based systems, has been a hot topic from the very beginning of the HTSC era,\cite{Dessau,Ma} and is still the subject of lively discussion today.\cite{Chuang,Fretwell,BorisPRL} In the past, the existence of a large, hole-like FS in angle-resolved photoemission spectroscopy (ARPES) was taken as support for the validity of Luttinger's theorem for the superconducting cuprates.\cite{Olson,Campuzano} While some ARPES studies of Bi-2212 conclude that a large, hole-like FS persists even to very low doping levels,\cite{Ding} other data imply a change in FS topology \cite{SchwallerEPJ} or the presence of hole-pockets at underdoping.\cite{Marshall} Recent data from La$_{2-x}$Sr$_x$CuO$_4$ (LSCO) have been interpreted in terms of a change of FS topology from hole-like for $x<0.2$ to electron-like for higher doping levels.\cite{Ino}

The recent improvement in the performance of photoemission instrumentation (in particular in the angular resolution) has led to a renaissance in the direct determination of the basal plane projection of the FS using ARPES. Considering the fundamental importance of the FS topology and shape in deciding the physical properties of a solid, it is natural to want to study its doping dependence in the Bi-based HTSC directly and with high precision using high resolution FS mapping.

The ARPES experiments reported here were performed either using monochromated He I radiation and an SES200 electron analyzer or using $h\nu$ = 25 eV radiation from the U125/1-PGM beamline at BESSY \cite{Follath} and an SES100 electron analyser. The samples were mounted on a triple-axis goniometer, enabling computer controlled angular scanning with a precision exceeding 0.1$^\circ$ for all axes, resulting in a dense sampling of a large portion of $\textbf{k}\omega$-space for each single crystal studied. The overall resolution was set to 0.014 \AA$^{-1} \times$ 0.035 \AA$^{-1} \times$ 19 meV which are the FWHM momentum (parallel and perpendicular to the analyzer entrance slit) and energy resolutions, respectively.\cite{BorisPRB} The samples were cleaved in-situ to give mirror-like surfaces and all data were measured above the pseudogapped regime at 300K within 3-4 hours of cleavage. The synchrotron based data were collected at 30~K with 0.014 \AA$^{-1} \times$ 0.014 \AA$^{-1} \times$ 17 meV resolution. We investigated a set of high quality single crystals of Pb-doped Bi-2212 (Pb:Bi ratio = 0.4:1.6) which had undergone different oxygen loading procedures. As we have pointed out earlier (see, e.g., Refs.~\onlinecite{BorisPRL,SibyllePRB,KordPRL}), it is wise to use the Pb-substituted variants for such experiments as these systems do not possess the incommensurate modulation of the BiO layers which in pristine Bi-2212 leads to the appearance of strong diffraction replicas of the main and shadow FS features in the maps, thus disqualifying a detailed discussion of the FS topology, shape and area as a function of doping. In the following, we label the samples, which span a $T_c$ range of 35 K around optimal doping, according to their $T_c$: UD 76K, UD 85K, UD 89K, OD 81K, OD 72K and OD 69K (UD and OD stand for underdoped and overdoped).

\begin{figure*}
\includegraphics[width=17.6cm]{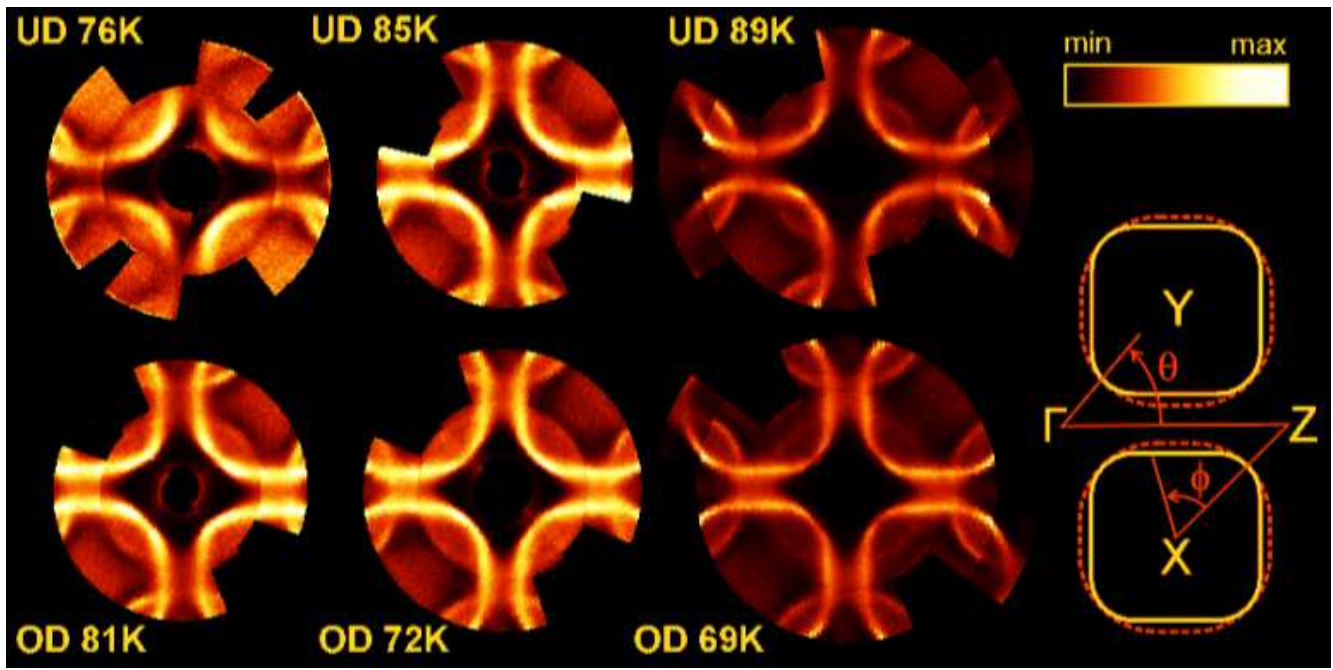}%
\caption{\label{maps} Basal plane projection of the normal state (300K) Fermi surface of Bi(Pb)-2212 from high resolution ARPES. The $E_F$ intensity (normalized to the signal at $\omega$ = 0.3 eV) is shown in color. The $T_c$ of each sample is indicated. The raw data cover half of the coloured area of each map, and have been rotated by 180$^\circ$ around the $\Gamma$ point to give a better k-space overview. The line dividing raw and rotated data runs almost vertically for the UD76K map, and from top left to bottom right in all other maps.The sketch shows the FS for the OD 69K dataset as yellow barrel-like shapes defined by joining the maxima of fits to the normalized $E_F$ MDC's.}
\end{figure*}

Fig.~\ref{maps} shows the Fermi surface maps for all six doping levels. Each dataset contains ca.~5000 ARPES spectra. 
We collect data from a significantly larger region of $\textbf{k}$-space than the irreducible octant, which brings the advantage of enabling a quantitative correction of angular misalignments of the crystal to a precision of 0.1$^\circ$.

To minimize the effects of the factors separating the ARPES intensity distribution from the spectral function (see Ref.~\onlinecite{BorisPRB}), the data were `self-normalized' by dividing the signal from the Fermi level, $I(\textbf{k},\omega=0)$, by the signal at highest binding energy, $I(\textbf{k},\omega_{hbe})$ (here $\omega_{hbe}$ is 300 meV). The FS topology and shape derived from these data do not depend sensitively upon the use of any reasonable self-normalization denominators, although we wish to stress here that the self-normalization procedure itself is indispensable for the precise determination of the \textbf{k}$_F$-vectors (see Appendix).

Before going on to discuss the data in a more quantitative manner, we first cover what can be learned directly from a simple visual inspection of Fig.~\ref{maps}. (i) There is no topological change of the main FS within the doping range studied - it remains hole-like (centered at the X, Y ($\pm \pi$,$\pm \pi$) points), in contrast to recent data from the LSCO system.\cite{Ino} (ii) As hole doping is increased, the main FS `barrels' increase in size (as can easily be seen in the decrease of the inter-barrel separation around the M ($\pi$,0) point), accompanied by an increase in the size of the lenses formed by main FS and shadow FS (SFS). (iii) The shape of the FS barrels changes from being quite rounded at low doping to take on the form of a square with well-rounded corners at higher doping. (iv) The SFS exists at all doping levels.

We stress that these statements describe experimental observations and are independent of any particular data analysis or physical interpretation.

One of the fundamental questions in the physics of 2D strongly correlated electron systems is to what extent the interacting electron system can be described by models derived perturbatively from the non-interacting case. One way to test this is to consider the validity or otherwise of Luttinger's theorem, which can be paraphrased by stating that the volume (area in 2D) of the FS should be conserved upon switching on the interactions. Thus if we are able to pin down the doping dependence of the exact path in $\textbf{k}$-space which represents the Fermi surface in, for example, the (Pb,Bi)-2212 HTSC without knowing, a priori, its shape, we would be able to evaluate the doping dependence of the FS area and thus test Luttinger's theorem. The best approach here is to locate the maxima in the $E_F$ momentum distribution curves \cite{Aebi} (MDC's) describing tracks crossing the FS (preferably at right angles, see Appendices here and in Ref.~\onlinecite{BorisPRB} for details).

Such a fitting procedure was carried out for the OD 69K sample. The detailed result is well described by a FS having the form of a square with rounded corners, which confirms the visual impression from the intensity map for this sample. Such a form gives a simple analytical approximation for the FS shape also predicted in tight-binding and LDA calculations. \cite{Liechtenstein,Andersen} A sketch of the fit result is shown as the yellow line on the right hand side of Fig.~\ref{maps}. The FS maps from the other samples were then fitted, whereby the extent of the straight sections, as well as the size of the barrel as a whole were varied to optimize the fit to the data. We can then derive the hole concentration $x$ from the simple relation $x + 1 = 2 S_b / S_{\textrm{BZ}}$, where $S_b$ is the area of main FS barrel and $S_{\textrm{BZ}}$ is the area of the Brillouin zone.

\begin{figure}
\includegraphics[width=8.6cm]{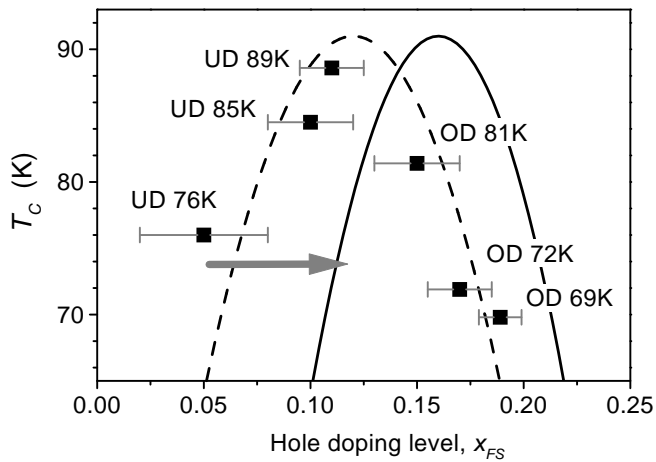}%
\caption{\label{doping} Symbols: critical temperatures vs the hole concentration, $x_{FS}$, the latter being calculated directly from the area of the FS's shown in Fig.~\ref{maps}. The solid line shows the commonly-used empirical relation for $T_c$ vs $x$ (Ref.~\onlinecite{Tallon}).}
\end{figure}

The results obtained from the analysis of the FS area are shown in Fig.~\ref{doping} in the form of a $T_c$ vs $x$ plot. The solid line shows the commonly employed empirical relation between $T_c$ and $x$.\cite{Tallon} For the six samples spanning a total of 35K in $T_c$, the co-ordinate pairs matching the $T_c$'s to the doping level taken directly from the experimentally determined Fermi surface area also give a parabolic curve (shown as a dotted line), but this curve is down-shifted in doping by ca.~0.05 towards the underdoped side of the phase diagram. This result, being quite surprising not long ago, can be well understood now in terms of the bilayer splitting of the CuO band.\cite{Feng,Chuang2} 

Before going further, we note that such a shift is hard to explain by the assumption that the doping level at the surface is lower than in the bulk. If this were the case (for example by loss of oxygen at the surface), such a deviation should be strongly dependent on the oxygen loading procedure, affecting the OD samples more strongly than the UD, which is clearly not the case. Furthermore, the fact that the superconducting gap seen in ARPES data from the same samples (not shown), closes unambiguously at the \textit{bulk} $T_c$ in the overdoped systems, is incompatible with a lower doping level at the surface.

\begin{figure}
\includegraphics[width=8.6cm]{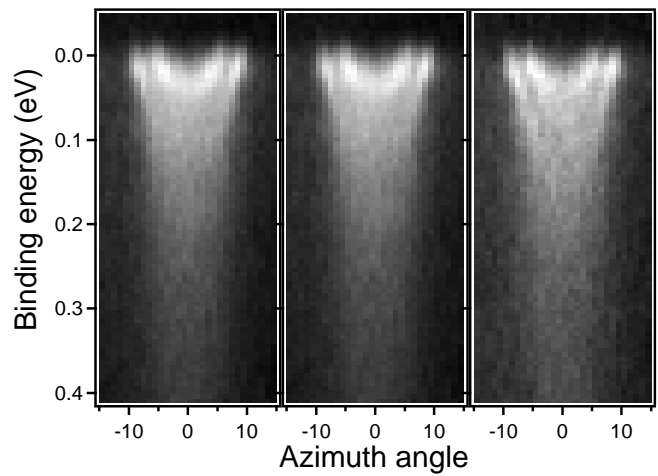}%
\caption{\label{split} The azimuthal energy distribution maps (EDM's) at $T$ = 30 K for three different $|\textbf{k}|$ = 1.084, 1.088 and 1.092 \AA$^{-1}$ (from the left to the right correspondingly) which demonstrate a well resolved bilayer splitting in underdoped (Bi,Pb)-2212 ($T_c$ = 77 K).}
\end{figure}

The acceptance of an existence of the $c$-axis bilayer splitting in Bi-2212 marks a watershed in the interpretation of ARPES data from the multilayer HTSC. This splitting has been directly resolved recently in highly overdoped Bi-2212 \cite{Feng,Chuang2} and (Bi,Pb)-2212 \cite{BogdanovPRB2001} and shown (but not resolved) to be roughly the same for underdoped Bi-2212 \cite{Chuang3}. In Fig.~\ref{split}, we show azimuthal energy distribution maps (EDM's: $I(\theta,\omega)$, where $\theta$ is the azimuth angle, see Fig.~\ref{maps}) at $T$ = 30 K for three different $|\textbf{k}|$ = 1.084, 1.088 and 1.092 \AA$^{-1}$ (from the left to the right correspondingly) which demonstrate a well resolved bilayer splitting in underdoped (Bi,Pb)-2212 ($T_c$ = 77 K).

Given the presence of the bilayer splitting (which we include here in the notion `complex structure'), the blurring of the FS (see Fig.~\ref{maps}) on going from the nodal to the antinodal point for all doping levels, which is often attributed to the complex physics of antinodal electrons (e.g. an absence of well-defined quasiparticles), could at least partially be due to such a complex structure of the FS itself. In order to examine this possibility, in the following we analyze the FS width in more detail.

\begin{figure}
\includegraphics[width=8.6cm]{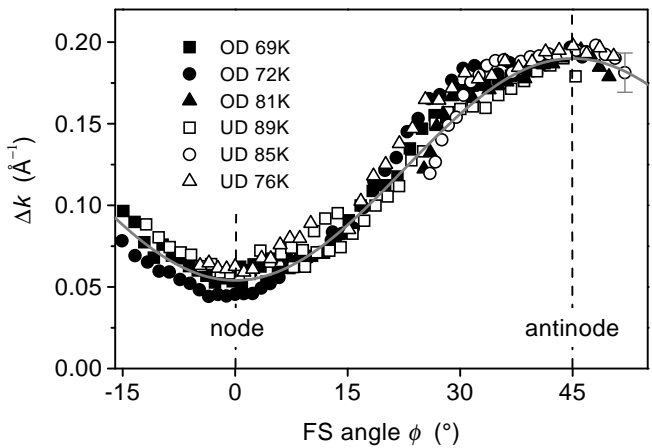}%
\caption{\label{width}The width of the room temperature main FS, $\Delta k$ vs the FS angle $\phi$ defined with respect to the nodal line. The $T_c$'s are indicated and the solid grey line represents the relation (\ref{E1}), for details see text.}
\end{figure}

In Fig.~\ref{width}, we show the width of the FS, $\Delta k$ vs $\phi$, the latter being the angle away from the nodal line, as indicated in Fig.~\ref{maps}. The $\Delta k$ values were derived from fitting $E_F$ MDC's using a Lorentzian profile with $\Delta k$ FWHM. For all doping levels investigated the room temperature FS width is strongly $k$-dependent, being maximal near the antinode and minimal at the node. The dotted line in Fig.~\ref{width} shows that the data can be well described by the function 
\begin{eqnarray}\label{E1}
\Delta k (\phi) = \Delta k_0 + \Delta k_1 \sin^2(2\phi), 
\end{eqnarray}
where $\Delta k_0$ = 0.054 \AA$^{-1}$ and $\Delta k_1$ = 0.136 \AA$^{-1}$. 

Remarkably, the observed $k$-dependence of the FS width is essentially \textit{independent} of the doping level. This is difficult to reconcile with a FS width determined solely by the complex physics of the FS electrons, as within such a picture the difference in the coupling to interactions between the nodal and antinodal regions should decrease continually as the doping increases. Equally, we can rule out effects resulting from differing group velocities around the FS contour, as these have been shown to be essentially constant.\cite{VallaPRL}

On the other hand, exploiting the `complex FS structure' scenario, we can associate a splitting in momentum, $\delta k (\phi)$, with the $c$-axis bilayer splitting. For the case in which the maxima of the MDCs (i.e.~the intensity in a self-normalized FS map such as those of Fig.~\ref{maps}) correspond to the inner bilayer split FS barrel (namely the bonding CuO-bilayer band), this would result in a shift of the observed doping level of 
\begin{eqnarray}\label{E2}
\delta x \approx \frac{\delta S}{S_{\textrm{BZ}}} \approx \frac{\left<k_b\right>}{S_{\textrm{BZ}}} \int_{0}^{2\pi} \delta k (\phi) d\phi,
\end{eqnarray}
where $\delta S$ is the difference in area between the split barrels, $k_b$ is the radius of main FS barrel with respect to the X-point ($\left<k_b\right> \approx$ 0.6$|\Gamma$X$|$) and $|\Gamma$X$|$ = 1.161 \AA$^{-1}$. This effect is illustrated schematically in the cartoon shown in Fig.~\ref{maps} where the yellow (red) barrels represent the smaller (larger) FS's resulting from the bilayer splitting. Taking a Lorentzian form for the $E_F$ MDC which cuts the FS, we expect a bilayer splitting induced FS width given by 
\begin{eqnarray}\label{E3}
\Delta k \approx W + \frac{3(\delta k)^2}{2 W},
\end{eqnarray}
where $W$ is the FWHM of the FS without splitting and $\delta k \le W / \sqrt{3}$ is assumed to hold). In such a manner we can estimate an upper limit for $\delta x$ = 0.07, which is illustrated in Fig.~\ref{doping} by the broad grey arrow. This demonstrates that the effect of the bilayer splitting is enough to explain the downshift of the $T_c$ vs doping parabola. 

The reason why the photocurrent intensity from the antibonding band at $E_F$ is less than from the bonding one and, consequently why the maxima of the MDCs correspond to the inner bilayer split FS barrel is the difference between matrix elements for photoemission from these two bands. In fact, the ratio between the effective matrix elements for emission from the bonding and antibonding Cu-O bands, $M_b/M_a \approx$ 2  for $h\nu$=21.2 eV (see Fig.~3 in Ref.~\onlinecite{KordPRL}). In contrast, for 25 eV excitation energy, $M_b/M_a \approx$ 1 and neither the emission from the bonding nor the antibonding band dominates resulting in a clear splitting as can be seen in the EDM's shown in Fig.~\ref{split}. 

Finally, we note in this context that the upper limits of $\delta k$ (and consequently $\delta x$) obtained above correspond to the limit at which two Lorentzian features are resolvable from one another: $\delta k = W / \sqrt{3}$. This same limit also defines the lower bound for the $\phi$-dependence of the Fermi surface width which arises from sources \textit{other} than the bilayer splitting: 
\begin{eqnarray}\label{E4}
\Delta k (\phi) = \Delta k_0 (1 + 1.3 \sin^2(2\phi)). 
\end{eqnarray}
In other words, this means that the detected anisotropy described by Eq.~(\ref{E1}) cannot be explained by the bilayer splitting alone. In considering either the `complex physics' or `complex FS structure' scenarios we discuss two extremes, whereas the real situation may well include contributions from both. For example, at high hole doping, the $\phi$-dependence of $\Delta k$ from `complex physics' should flatten out, which would be counteracted by the increasing bilayer splitting for this doping regime (in which the flat bands approach closer to $E_F$). Conversely, at low hole doping, the $\phi$-dependence of the coupling to interactions is strong, whereas the bilayer splitting would be expected to be weaker.\cite{LindroosXXX} In this way we end up with the observed overall doping independence of $\Delta k(\phi)$.

As mentioned above, it is possible to compensate for the downshift of the $T_c$ vs. $x$ parabola in Fig.~2 by taking the bilayer splitting into account. It would then follow that the area of the main ARPES FS scales with (1+$x$) in holes across the complete doping range studied. This behaviour is in contrast to what is seen in transport measurements. Resistivity and Hall effect data indicate that the transport characteristics scale with $x$,\cite{Batlogg,Ong} even into the overdoped regime.\cite{Alloul} Although it is conceivable that only those mobile electrons which have relatively low coupling to other degrees of freedom contribute to the transport, it is surely more than coincidental that this proportion should be exactly $x$/(1+$x$). This fundamental difference between the transport data and the ARPES FS is a key question which deserves detailed theoretical attention.

\begin{figure}
\vspace{0.2cm}
\includegraphics[width=8cm]{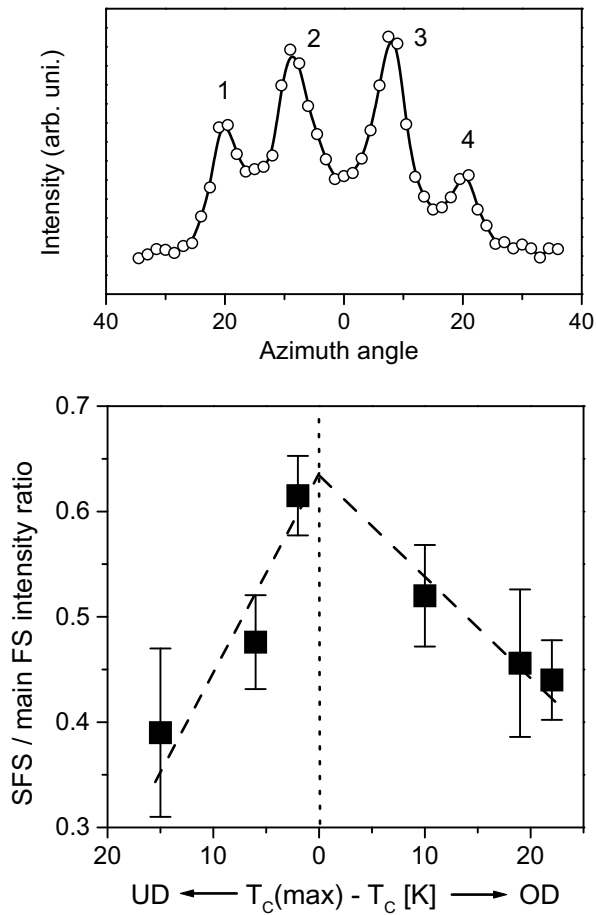}%
\caption{\label{shadow} SFS to main FS intensity ratios vs $T_c^{max} - T_c$ (lower panel), the dashed straight lines are guides to the eye; and an example of an azhimuth MDC from which these intensity ratios have been determined (upper panel): peaks 1 and 4 correspond to the SFS, peaks 2 and 3 correspond to the main FS, then the intensity ratio SFS/FS $= (I_1+I_4)/(I_2+I_3)$.}
\end{figure}

A final surprise that the FS has in store for us is shown in Fig.~\ref{shadow} (lower panel), in which the doping dependence of the intensity ratio of the SFS to that of the main FS is plotted. The intensities $I_m$ were taken in each case from the same azimuthal MDC scan: i.e.~with the same $|\textbf{k}|$ value, some 0.13 \AA$^{- 1}$ from the point at which the SFS and main FS `cross'.  For this scan the given intensity ratio reaches a local maximum as a function of $|\textbf{k}|$ which is a consequence of different dependencies of the photocurrent from the main and shadow FS's (at the same $|\textbf{k}|$) on matrix elements. The upper panel of Fig.~\ref{shadow} shows an example (for the OD 69K) of such azimuth MDCs from which these intensity ratios have been determined: peaks 1 and 4 correspond to the SFS, peaks 2 and 3 correspond to the main FS, then the intensity ratio SFS/FS $= (I_1+I_4)/(I_2+I_3)$. As Fig.~\ref{shadow} shows, this ratio decreases not only on going from optimal to overdoping, but also on going towards the underdoped side of the phase diagram (the rate of change is, in fact, even faster on the UD side). This is in contrast to predictions based on an antiferromagnetic origin of the SFS.\cite{Haas} The fact that the relative strength of the SFS tracks the doping dependence of T$_C$ means that, regardless of whether the SFS has structural or other origins, this phenomenon is important and could be related to high $T_c$ superconductivity itself. The behaviour seen here, taken together with the very strong similarities with SFS data from pristine Bi2212 (which has important \textit{structural} differences to Pb-doped Bi2212)  means that further work is needed, both on the experimental but also on the theoretical side, before the question of the origin and consequences of the shadow Fermi surface can be considered as being solved.

In conclusion, we have presented a detailed and systematic ARPES investigation of the doping dependence of the normal state (room temperature) FS of the Bi-2212 family of HTSC materials. The data clearly show no change in the FS topology away from hole-like at any stage (from UD 76K to OD 69K). An analysis of the main FS area gives a parabolic $T_c$ vs $x_{FS}$ relation, shifted to lower $x$ by some 0.05 compared to the `universal' relation,\cite{Tallon} which can be accounted for by the presence of two (unresolved) FS's near ($\pi$,0) due to a bilayer splitting with a maximum value ca.~0.05 \AA$^{-1}$, which stays roughly constant across the whole doping range. Furthermore, the FS width is shown to be strongly dependent on $k$, but for each particular $k_F$ point it is essentially independent of the doping level. This can be understood as a combination of the effects of the bilayer splitting (dominating at higher doping) and the complex physics of the FS electrons (dominating at lower doping). Finally, the shadow FS is clearly visible for all doping levels, and has maximal intensity at optimal doping, raising the question of a possible link between the origins of the shadow FS and superconductivity.

We are grateful to the BMBF (05 SB8BDA 6) and to the Fonds National Suisse de la Recherche Scientifique for support and to S.-L.~Drechsler, A.~N.~Yaresko, A.~Ya.~Perlov, R.~Hayn, N.~M.~Plakida, M.~Eschrig, and O.~K.~Andersen for stimulating discussions. 

\appendix*
\section{}

A meaningful estimation of the doping level from a FS map as described above requires the very precise determination of the FS vectors. Here we explain why the self-normalization procedure has been chosen for this purpose.

\begin{figure}[b!]
\vspace{0.2cm}
\includegraphics[width=7.4cm]{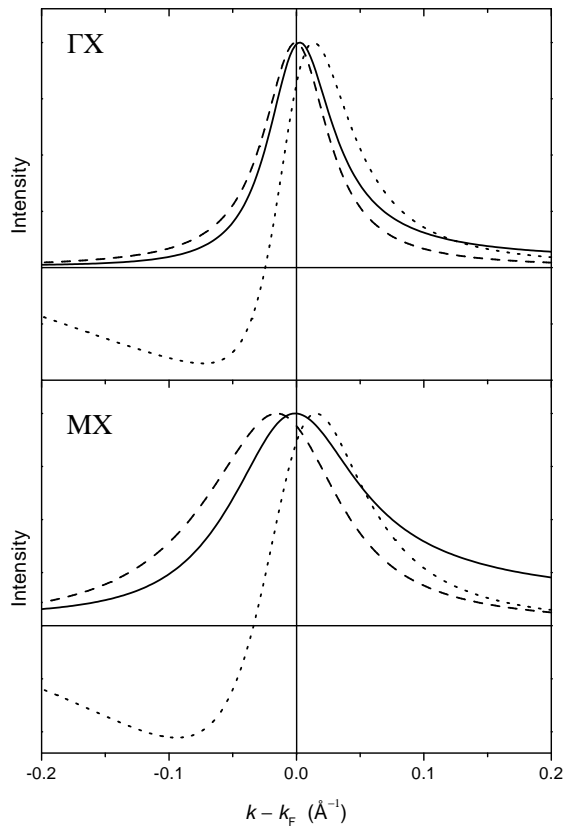}%
\caption{\label{norms} Results of $E_F$-MDC simulations for nodal ($\Gamma-(\pi,\pi)$ crossing, upper panel) and antinodal ($(\pi,0),(\pi,\pi)$ crossing, lower panel) ARPES data for a typical low-energy dispersion (for details see text). The dashed curves are the raw (non-normalized) MDCs, the solid curves are the MDCs after self-normalization to the highest binding energy, and the dotted curves represent the results of the $dn(k)/dk$ method.}
\end{figure}

In Ref.~\onlinecite{BorisPRB} we already discussed the applicability of different methods of $k_F$ determination to ARPES data from Bi-cuprates and demonstrated that the most accurate is the `maximum MDC' method.\cite{Aebi} In Ref. ~\onlinecite{BorisPRB} this was illustrated for the case of the nodal direction. At lower energy resolution (or for the case in which the MDC peaks are broader), however, the deviation of the experimentally determined $k_F$ from the true value, $\Delta k_F$, could be considerable, and even comparable with the $\Delta k_F$'s from other methods such as `gradient $n(k)$' (see Ref.~\onlinecite{BorisPRB}). It turns out that the above mentioned shift ($\Delta k_F$) is nearly completely compensated for by the self-normalization procedure and therefore the deviation of the visible FS traces on the self-normalized intensity maps (like those shown in Fig.~\ref{maps}) from the real FS is negligible. This is demonstrated below. 

Fig.~\ref{norms} represents the results of $E_F$-MDC simulations in the nodal ($\Gamma-(\pi,\pi)$ crossing, upper panel) and antinodal (($(\pi,0)-(\pi,\pi)$ crossing, lower panel) points for typical low-energy dispersion relations $\varepsilon_k = v_F(k_F-k)$ with $v_F = 2$ eV\AA. For this simulation we use a simple form for the spectral function (with the momentum resolution included)
\begin{eqnarray}\label{A1}
A(k,\omega,R_{k}) \propto \frac{\sqrt{\Sigma''(\omega,T)^{2} + R_{k}^{2}}}{(\omega - \varepsilon_{k})^{2} + \Sigma''(\omega,T)^{2} + R_{k}^{2}}, 
\end{eqnarray}
where the imaginary part of the self-energy 
\begin{eqnarray}\label{A2}
\Sigma''(\omega,T) = \sqrt{(\alpha \omega)^2 + (\beta T)^2}
\end{eqnarray}
has been taken with $\alpha = 1$ and $\beta = 2$ (binding energy $\omega$ and temperature $T=300$K are in energy units), giving a resonable fit to the experiment \cite{BorisPRB,KordPRL}. The real part of the self-energy is included in $\varepsilon_k$. The photocurrent is calculated as 
\begin{eqnarray}\label{A3}
I(k,\omega) \propto [A(k,\omega,R_{k}) f(\omega)] \otimes R_{\omega},
\end{eqnarray}
where $f(\omega)$ is the Fermi function. The momentum resolutions are taken to be 0.014 \AA$^{-1}$ and 0.035 \AA$^{-1}$ for the $\Gamma$X and MX crossings respectively and energy resolution of 19 meV for both. For the MX crossing, to check an extreme, the `splitting value' $\Delta\varepsilon=80$ meV is added to the $R_{\omega}$ function as a FWHM. The dashed curves represent the raw (non-normalized) MDCs: $I(k,0)$. The solid curves represent MDCs obtained by self-normalizing every EDC to the highest binding energy ($\omega_{hbe}$ in this case is 0.3 eV): $I(k,0)/I(k,\omega_{hbe})$. For comparison, the $dn(k)/dk$ dependencies are shown as dotted lines where 
\begin{eqnarray}\label{A4}
n(k) = \int_{-\omega_{hbe}}^{\omega_{hbe}}I(k,\omega) d\omega .
\end{eqnarray}

Fig.~\ref{norms} illustrates that whereas in the case of the nodal region both raw and self-normalized MDCs are only slightly shifted from the real $k_F$ ($\Delta k_F = -0.001$ and 0.003 \AA$^{-1}$ respectively), in the antinodal region the shift of the raw MDC is rather large ($-0.015$ \AA$^{-1}$) whereas the peak of the self-normalized curve practically coincides with the true $k_F$ (i.e. $\Delta k_F = -0.001$ \AA$^{-1}$). This demonstrates the power of the self-normalization procedure: its application to the intensity maps not only reduces the influence of the matrix element effects \cite{BorisPRB} but also restores the true 
location of the FS vectors, thus making it the correct choice in the study of the FS topology, shape and area.


\begin{thebibliography}{}
\bibitem{Dessau} D. S. Dessau, Z.-X. Shen, D. M. King, D. S. Marshall, L. W. Lombardo, P. H. Dickinson, A. G. Loeser, J. DiCarlo, C.–H Park, A. Kapitulnik, and W. E. Spicer, Phys. Rev. Lett. \textbf{71}, 2781 (1993).
\bibitem{Ma} Jian Ma, C. Quitmann, R. J. Kelley, P. Alm\'eras, H. Berger, G. Margaritondo, and M. Onellion, Phys. Rev. B \textbf{51}, 3832 (1995).
\bibitem{Chuang} Y.-D. Chuang, A. D. Gromko, D. S. Dessau, Y. Aiura, Y. Yamaguchi, K. Oka, A. J. Arko, J. Joyce, H. Eisaki, S. I. Uchida, K. Nakamura, and Yoichi Ando, Phys. Rev. Lett. \textbf{83}, 3717 (1999).
\bibitem{Fretwell} H. M. Fretwell, A. Kaminski, J. Mesot, J. C. Campuzano, M. R. Norman, M. Randeria, T. Sato, R. Gatt, T. Takahashi, and K. Kadowaki, Phys. Rev. Lett. \textbf{84}, 4449 (2000).
\bibitem{BorisPRL} S. V. Borisenko, M. S. Golden, S. Legner, T. Pichler, C. D\"urr, M. Knupfer, J. Fink, G. Yang, S. Abell, and H. Berger, Phys. Rev. Lett. \textbf{84}, 4453 (2000).
\bibitem{Olson} C. G. Olson, R. Liu, D. W. Lynch, R. S. List, A. J. Arko, B. W. Veal, Y. C. Chang, P. Z. Jiang, and A. P. Paulikas, Phys. Rev. B \textbf{42}, 381 (1990).
\bibitem{Campuzano} J. C. Campuzano, G. Jennings, M. Faiz, L. Beaulaigue, B. W. Veal, J. Z. Liu, A. P. Paulikas, K. Vandervoort, H. Claus, R. S. List, A. J. Arko, and R. J. Bartlett, Phys. Rev. Lett. \textbf{64}, 2308 (1990).
\bibitem{Ding} H. Ding, M. R. Norman, T. Yokoya, T. Takeuchi, M. Randeria, J. C. Campuzano, T. Takahashi, T. Mochiku, and K. Kadowaki, Phys. Rev. Lett. \textbf{78}, 2628 (1997).
\bibitem{SchwallerEPJ} P. Schwaller, T. Greber, P. Aebi, J.M. Singer, H. Berger, L. F\'orro, J. Osterwalder, Eur. Phys. J. B \textbf{18}, 215 (2000).
\bibitem{Marshall} D. S. Marshall, D. S. Dessau, A. G. Loeser, C-H. Park, A. Y. Matsuura, J. N. Eckstein, I. Bozovic, P. Fournier, A. Kapitulnik, W. E. Spicer, and Z.-X. Shen, Phys. Rev. Lett. \textbf{76}, 4841 (1996).
\bibitem{Ino} A. Ino, C. Kim, M. Nakamura, T. Yoshida, T. Mizokawa, Z.-X. Shen, A. Fujimori, T. Kakeshita, H. Eisaki, and S. Uchida, cond-mat/0005370.
\bibitem{Follath} R. Follath, Nucl. Instr. and Meth. A \textbf{467-468}, 418 (2001).
\bibitem{BorisPRB} S. V. Borisenko, A. A. Kordyuk, S. Legner, C. D\"urr, M. Knupfer, M. S. Golden, J. Fink, K. Nenkov, D. Eckert, G. Yang, S. Abell, H. Berger, L. Forr\'o, B. Liang, A. Maljuk, C. T. Lin, and B. Keimer, Phys. Rev. B \textbf{64}, 094513 (2001).
\bibitem{SibyllePRB} S. Legner, S. V. Borisenko, C. D\"urr, T. Pichler, M. Knupfer, M. S. Golden, J. Fink, G. Yang, S. Abell, H. Berger, R. M\"uller, C. Janowitz, and G. Reichardt, Phys. Rev. B \textbf{62}, 154 (2000).
\bibitem{KordPRL} A. A. Kordyuk, S. V. Borisenko, T. K. Kim, K. Nenkov, M. Knupfer, M. S. Golden, J. Fink, H. Berger, R. Follath, cond-mat/0110379.
\bibitem{Aebi} P. Aebi, J. Osterwalder, P. Schwaller, L. Schlapbach, M. Shimoda, T. Mochiku, and K. Kadowaki, Phys. Rev. Lett. \textbf{72}, 2757 (1994).
\bibitem{Liechtenstein} A. I. Liechtenstein, O. Gunnarsson, O. K. Andersen, and R. M. Martin, Phys. Rev. B \textbf{54}, 12505 (1996).
\bibitem{Andersen} O. K. Andersen, A. I. Liechtenstein, O. Jepsen, and F. Paulsen, J. Phys. Chem. Solids \textbf{56}, 1573 (1995).
\bibitem{Tallon} J. L. Tallon, C. Bernhard, H. Shaked, R. L. Hitterman, and J. D. Jorgensen, Phys. Rev. B \textbf{51}, 12911 (1995)
\bibitem{Feng} D. L. Feng, N. P. Armitage, D. H. Lu, A. Damascelli, J. P. Hu, P. Bogdanov, A. Lanzara, F. Ronning, K. M. Shen, H. Eisaki, C. Kim, Z.-X. Shen, J.-i. Shimoyama and K. Kishio, Phys. Rev. Lett. \textbf{86}, 5550 (2001).
\bibitem{Chuang2} Y.-D. Chuang, A. D. Gromko, A. Fedorov, Y. Aiura, K. Oka, Yoichi Ando, H. Eisaki, S. I. Uchida, and D. S. Dessau, Phys. Rev. Lett. \textbf{87}, 117002 (2001).
\bibitem{BogdanovPRB2001} P. V. Bogdanov, A. Lanzara, X. J. Zhou, S. A. Kellar, D. L. Feng, E. D. Lu, H. Eisaki, J.-I. Shimoyama, K. Kishio, Z. Hussain, and Z.-X. Shen, Phys. Rev. B \textbf{64}, 180505(R) (2001).
\bibitem{Chuang3} Y.-D. Chuang, A. D. Gromko, A. V. Fedorov, Y. Aiura, K. Oka, Yoichi Ando, D. S. Dessau, cond-mat/0107002.
\bibitem{Krakauer} H. Krakauer and W. E. Pickett, Phys. Rev. Lett. \textbf{60}, 1665 (1988).
\bibitem{Massida} S. Massida, J. Yu, and A. J. Freeman, Physica C \textbf{152}, 251 (1988).
\bibitem{VallaPRL} T. Valla, A. V. Fedorov, P. D. Johnson, Q. Li, G. D. Gu, and N. Koshizuka, Phys. Rev. Lett. \textbf{85}, 828 (2000).
\bibitem{LindroosXXX} M. Lindroos, S. Sahrakorpi, A. Bansil, cond-mat/0109039.
\bibitem{Batlogg} H. Takagi, B. Batlogg, H. L. Kao, J. Kwo, R. J. Cava, J. J. Krajewski, and W. F. Peck, Phys. Rev. Lett., \textbf{69}, 2975 (1992).
\bibitem{Ong} N. P. Ong in \textit{Physical Properties of High- Temperature Superconductors}, edited by D. M. Ginzberg (World Scientific Singapore, 1990) Vol. 2, p 459.
\bibitem{Alloul} F. Rullier-Albenque, P. A. Vieillefond, H. Alloul, A. W. Tyler, P. Lejay, and J. F. Marucco, Europhys. Lett. \textbf{50}, 81 (2000).
\bibitem{Haas} S. Haas, A. Moreo, and E. Dagotto, Phys. Rev. Lett. \textbf{74}, 4281 (1995).
\end{thebibliography}
\end{document}